\documentclass[11pt,a4paper]{article}
\AtBeginDocument{}
\usepackage{hyperref}
\usepackage{amsfonts,amssymb,amsmath}
\usepackage{soul}

\def\G{\Gamma}

\begin{document}
\
\vskip 1 truecm
{\Large \bf
\centerline{Quantum Local Symmetry of the $D$-Dimensional}
\centerline{Non-Linear Sigma Model: A Functional Approach}
}

\vskip 0.7 truecm
\centerline{Andrea Quadri $^{a,b}$\footnote{e-mail address: \tt andrea.quadri@mi.infn.it}}

\vskip 0.3 truecm
\begin{center}
$^{a}$ Istituto Nazionale di Fisica Nucleare (INFN), Sezione di Milano,\\ via Celoria 16, I-20133 Milano, Italy
\\
$^{b}$ Dipartimento di Fisica, Universit\`a di Milano, via Celoria 16, I-20133 Milano, Italy

\end{center}

\vskip 0.4 truecm
\centerline{\bf Abstract}

\begin{quotation}We summarize recent progress on the symmetric subtraction of the Non-Linear Sigma Model in $D$ dimensions, based on the validity of a certain Local Functional Equation (LFE) encoding the invariance of the SU(2) Haar measure under local left transformations. The deformation of the classical non-linearly realized symmetry at the quantum level is analyzed by cohomological tools. It is shown that all the divergences of the one-particle irreducible (1-PI) amplitudes (both on-shell and off-shell) can be classified according to the solutions of the LFE. Applications to the non-linearly realized Yang-Mills theory and to the electroweak theory, which is directly relevant to the model-independent analysis of LHC data, are briefly addressed.
\end{quotation}


\vskip 0.8 truecm
Invited review published in Symmetry {\bf 6}(2), 234-255.

\newpage

\vspace {-12pt}
\section{Introduction}

The purpose of this paper is to provide an introduction to the recent advances
in the study of the renormalization properties
of the SU(2) Non-Linear Sigma Model (NLSM) and of the quantum
deformation of the underlying non-linearly realized
classical SU(2) local symmetry.
The results reviewed here are based mainly on
Refs.~\cite{Ferrari:2005ii}-\cite{Ferrari:2011bx}.

The linear sigma model was originally proposed a long time ago
in~\cite{GellMann:1960np} in the context of
elementary particle physics. In this model
the pseudoscalar pion fields $\vec{\phi}$ form a chiral
multiplet together with a scalar field $\sigma$,
with ($\sigma$, $\vec{\phi}$) transforming linearly
as a vector under $\rm O(4) \sim SU(2)\times SU(2) / Z_2$.
If one considers instead the model on the manifold defined by
\begin{eqnarray}
\sigma^2 + \vec{\phi}^2 = f_\pi^2 \, , ~~~~~ \sigma > 0 \,
\end{eqnarray}
one obtains a theory where the chiral group $\rm SO(4) \sim SU(2)\times SU(2)$
(with $\rm SO(4)$ selected by the positivity condition on $\sigma$)
is spontaneously broken down to the isotopic spin group $\rm SU(2)$.
The composite field $\sigma$ has a non-vanishing expectation value $f_\pi$ (to be identified with the pion decay constant), while the pions are massless.
Despite the fact that this is only an approximate description (since in reality the pions are massive and chiral $\rm SU(2) \times SU(2)$ is not exact, even
before being spontaneously broken), the approach turned out
to be phenomenologically quite successful  and paved the way to
the systematic use of effective field theories as a low energy expansion.

The first step in this direction was to obtain a phenomenological
lagrangian directly, by making use of a pion
field with non-linear transformation properties
dictated by chiral symmetry from the beginning.
After the seminal work of Reference~\cite{Weinberg:1968de} for the
chiral $\rm SU(2) \times SU(2)$ group,
non-linearly realized symmetries
were soon generalized to arbitrary groups
in~\cite{Coleman:1969sm,Callan:1969sn} and have since then become
a very popular tool~\cite{Weinberg:1978kz}.

Modern applications involve, e.g., Chiral Perturbation
Theory~\cite{Gasser:1983yg}-\cite{Ecker:1989yg},
low energy electroweak theories~\cite{Buchmuller:1985jz} as well as gravity~\cite{Donoghue:1995cz}.

Effective field theories usually exhibit an infinite number
of interaction terms, that can be organized according to the
increasing number of derivatives.
By dimensional arguments,
the interaction terms must then be suppressed
by some large mass scale $\rm M$ (so that one expects that the
theory is reliable at energies well below $\rm M$) (For a modern introduction to the problem, see e.g.,~\cite{Weinberg:1996kr}).
In the spirit of the phenomenological lagrangians, the tree-level effective action is used to compute physical quantities up to a given order
in the momentum expansion. Only a finite number of
derivative interaction vertices contribute to that order,
thus allowing to express the physical observables one is interested in
through a finite number of parameters (to be eventually fixed by comparison
with experimental data).
Then the theory can be used to make predictions at the given order of accuracy
in the low-energy expansion.


The problem of the mathematically consistent evaluation of quantum
corrections in this class of models has a very long history.
On general grounds, the derivative couplings
tend to worsen the ultraviolet (UV) behavior
of the theory, since
UV divergent contributions arise in the
Feynman amplitudes that cannot be compensated
by a multiplicative renormalization of the fields
and a
redefinition of the mass parameters and the coupling constants
in the classical action (truncated at some
given order in the momentum expansion).
Under these circumstances, one says that the theory is \linebreak non-renormalizable (A compact introduction to renormalization theory is given in~\cite{Itzykson:1980rh}).

It should be stressed that the key point here is the instability
of the classical action: no matter how many terms are kept in the derivative expansion of the tree-level action, there exists a sufficiently high loop order
where UV divergences appear that cannot be reabsorbed into the classical action. On the other hand, if in a non-anomalous and non-renormalizable gauge theory
 one allows for
{\em infinitely many} terms in the classical action (all those compatible with
the symmetries of the theory), then UV divergences can indeed
be reabsorbed by preserving the Batalin-Vilkovisky master
equation~\cite{Gomis:1994he}  and the model is said to be renormalizable in the modern sense~\cite{Gomis:1995jp}.

Sometimes symmetries are so
powerful in constraining the UV divergences
that the non-linear theory proves to be indeed renormalizable
(although not by power-counting), like for instance
the NLSM in two dimensions
\cite{Brezin:1976ap,Becchi:1988nh} (For a more recent introduction to the subject, see e.g. \cite{ZinnJustin:2002ru}).


In four dimensions the situation is much less favorable.
It has been found many years ago that already
at one loop level in the four-dimensional NLSM
there exists an infinite number of
one-particle irreducible (1-PI)
divergent pion amplitudes.
Many attempts were then made in the literature in order to classify such divergent terms.
Global SU(2) chiral symmetry is not preserved already at one loop level
~\cite{Ecker:1972bm}-\cite{Tataru:1975ys}.
Moreover it turns out that some of the non-symmetric terms can be reabsorbed by a redefinition of the 
fields~\cite{Tataru:1975ys}-\cite{Honerkamp:1996va},
%
 however  in the off-shell four-point $\phi_a$ amplitudes some divergent parts arise that cannot be reabsorbed by field redefinitions unless derivatives are allowed~\cite{Tataru:1975ys}.
These technical difficulties prevented such attempts to evolve into a mathematically consistent subtraction procedure.

More recently it has been pointed out~\cite{Ferrari:2005ii} that
one can get the full control  on the ultraviolet
divergences of the $\phi$'s-amplitudes by exploiting
 the constraints stemming from the presence
of a certain local symmetry, associated with the
introduction of a SU(2) background field connection
into the theory. This symmetry in encoded in functional form
in the so-called Local Functional Equation (LFE)~\cite{Ferrari:2005ii}.
It turns out that the fundamental divergent amplitudes
are not those associated with the quantum fields of the theory,
namely the pions, but those corresponding to the background connection
and to the composite operator implementing the non-linear constraint~\cite{Ferrari:2005ii,Ferrari:2005va}. These amplitudes are named ancestor amplitudes.

At every order in the loop expansion
there is only a finite number of divergent ancestor amplitudes.
They uniquely fix the divergent amplitudes involving the pions.
Moreover, non-renormalizability of this theory
in four dimensions can be traced back to the instability of the classical
non-linear local symmetry, that gets
deformed by quantum corrections. These results hold for the
full off-shell amplitudes~\cite{Bettinelli:2007kc}.

A comment is in order here. In Reference~\cite{Bettinelli:2007zn}
it has been argued that Minimal Subtraction is a symmetric scheme,
fulfilling all the symmetries of the NLSM in the LFE approach.
This in particular entails that all finite parts of the needed
higher order counterterms are consistently set to zero. It should
be stressed that this is not the most general solution compatible
with the symmetries and the WPC, that is commonly used in the
spirit of the most popular effective field theory point of view.
Indeed, these finite parts are constrained neither by the LFE
nor by the WPC  and thus, mathematically, they can be freely chosen, as far as they are introduced at the order prescribed by the WPC and without violating the~LFE.

The four dimensional SU(2) NLSM
provides a relatively simple playground where to test
the approach based on the LFE,
 that can be further generalized to the SU(N) case (and
possibly even to a more general Lie group).

Moreover,
when the background vector field becomes dynamical, the SU(2) NLSM action allows one to generate a mass term for the
gauge field {\em \`a la} St\"uckelberg~\cite{Stueckelberg:1938zz,Ruegg:2003ps}.
The resulting non-linear implementation of the spontaneous
symmetry breaking mechanism (as opposed to the linear
Higgs mechanism) is widely used in the context of
electroweak low energy effective field theories, that are
a very important tool in the  model-independent analysis of
LHC data~\cite{Altarelli:2000ye}-\cite{Espinosa:2012im}.

\section{The Classical Non-Linear Sigma Model}

The classical SU(2) NLSM
in $D$ dimensions is defined by the action
\begin{equation}
S_0 = \int d^Dx\, \frac{m_D^2}{4} {\rm Tr} \Big ( \partial_\mu \Omega^\dagger
\partial^\mu \Omega \Big ) \,
\label{cl.act}
\end{equation}
where the matrix $\Omega$ is a SU(2) group element  given by
\begin{equation}
\Omega = \frac{1}{m_D} ( \phi_0 + i \phi_a \tau_a ) \, ,
\quad
\Omega^\dagger \Omega = 1 \, ,
\quad
{\rm det} ~ \Omega  = 1 \, ,
\quad
\phi_0^2 + \phi_a^2 = m_D^2 \,
\label{fields}
\end{equation}
In the above equation $\tau_a$, $a=1,2,3$ are the Pauli matrices and
$m_D = m^{D/2-1}$ is the mass scale of the theory. $m$ has mass dimension $1$.
$\phi_a$ are the three independent fields parameterizing the matrix $\Omega$,
while we choose the positive solution of the non-linear constraint,
yielding
\begin{equation}
\phi_0 = \sqrt{m_D^2 - \phi_a^2} \,
\label{nl.constraint}
\end{equation}
In components one finds
\begin{eqnarray}
S_0 =  \int d^Dx\, \Big (
\frac{1}{2} \partial_\mu \phi_a \partial^\mu \phi_a
+ \frac{1}{2} \frac{\phi_a \partial_\mu \phi_a \phi_b \partial^\mu \phi_b}{\phi_0^2} \Big ) \,
\label{cl.act.comp}
\end{eqnarray}
The model therefore contains non-polynomial, derivative interactions for the
massless scalars $\phi_a$.

Equation (\ref{cl.act}) is invariant under a {\em global} $\rm SU(2)_L \times SU(2)_R$ chiral transformation
\begin{eqnarray}
\Omega' = U \Omega V^\dagger\, , \quad U \in {\rm SU(2)_L} \, , ~~
V \in {\rm SU(2)_R} \,
\label{global.su2}
\end{eqnarray}
%

We notice that such a global transformation is non-linearly realized,
as can be easily seen by looking at its infinitesimal version.
E.g., for the left transformation one finds:
\begin{eqnarray}
\delta \phi_a = \frac{1}{2} \alpha \phi_0(x) + \frac{1}{2} \epsilon_{abc} \phi_b(x) \alpha_c \, , \qquad \delta \phi_0(x) = -\frac{1}{2} \alpha \phi_a(x) \,
\label{global.su2.comp}
\end{eqnarray}
Since $\phi_0$ is given by Equation (\ref{nl.constraint}), the first term
in the r.h.s. of $\delta \phi_a$ is non-linear (and even non-polynomial) in the quantum fields.

Perturbative quantization of the NLSM requires to carry out the
path-integral
\begin{eqnarray}
Z[J] = \int {\cal D} \phi_a \,
\exp \Big ( i S_0[\phi_a] + i \int d^Dx \, J_a \phi_a \Big ) \,
\label{path.integral}
\end{eqnarray}
by expanding around the free theory and by treating the second term
in the r.h.s.~of Equation (\ref{cl.act.comp}) as an interaction.
Notice that in Equation (\ref{path.integral}) the sources $J_a$ are coupled
to the fields $\phi_a$ over which the path-integral is performed.
In momentum space the propagator for the $\phi_a$ fields is
\begin{eqnarray}
\Delta_{\phi_a \phi_b} = i \frac{\delta_{ab}}{p^2} \,
\label{prop}
\end{eqnarray}
The mass dimension of the $\phi_a$ is therefore $D/2-1$, in agreement
with Equation (\ref{fields}).

The presence of two derivatives in the interaction term is the cause (in dimensions
greater than 2) of severe UV divergences, leading to the non-renormalizability
of the theory.

\section{The Approach based on the Local Functional Equation}


Some years ago it was recognized that the most effective classification of the UV divergences
(both for on-shell and off-shell amplitudes) of the NLSM cannot be achieved in terms of the
quantized fields $\phi_a$, as it usually happens in power-counting renormalizable theories, but rather through the
so-called ancestor amplitudes, \emph{i.e.}, the Green's functions of certain composite operators, whose knowledge
completely determines the amplitudes involving at least one
$\phi_a$-leg.
This property follows as a consequence of the existence
of an additional local functional identity, the so-called Local Functional Equation (LFE)~\cite{Ferrari:2005ii}.

The LFE  stems from the {\em local} $\rm SU(2)_L$-symmetry that can be established
from the gauge transformation of the flat connection $F_\mu$ associated with
the matrix $\Omega$:
\begin{eqnarray}
F_\mu = i \Omega \partial_\mu \Omega^\dagger = \frac{1}{2} F_{a\mu} \tau^a \,
\label{flat.conn}
\end{eqnarray}
\emph{i.e.}, the  local $\rm SU_L(2)$-transformation of $\Omega$


%
\begin{eqnarray}
\Omega ' = U \Omega \,
\label{su2.local.omega}
\end{eqnarray}
induces a gauge transformation of the flat connection, namely
\begin{eqnarray}
F'_\mu = U F_\mu U^\dagger+ i U \partial_\mu U^\dagger \,
\label{su2.local.flatconn}
\end{eqnarray}
$S_0$ in Equation (\ref{cl.act})  is not invariant under local
$\rm SU(2)_L$ transformations; however it is easy to made
it invariant, once one realizes that it can be written as
\begin{eqnarray}
S_0 = \int d^Dx \, \frac{m_D^2}{4} Tr (F_\mu^2) \,
\label{cl.act.flat}
\end{eqnarray}
Since $F_\mu$ transforms as a gauge connection, one can
introduce an additional external classical vector
source $\tilde J_{\mu} = \frac{1}{2} \tilde J_{a\mu} \tau^a$ and
replace $S_0$ with
\begin{eqnarray}
S = \int d^Dx \,  \frac{m_D^2}{4} Tr ( F_\mu - \tilde J_\mu)^2 \,
\label{cl.act.lfe}
\end{eqnarray}
If one requires that $\tilde J_{a\mu}$ transforms as a gauge
connection under the local $\rm SU(2)_L$ group, $S$ in
Equation~(\ref{cl.act.lfe}) is invariant under a local
$\rm SU(2)_L$ symmetry given by
\begin{eqnarray}
&& \delta \phi_a = \frac{1}{2} \alpha_a \phi_0 + \frac{1}{2} \epsilon_{abc} \phi_b \alpha_c \, , \qquad \delta \phi_0 = -\frac{1}{2} \alpha_a \phi_a \, 
\nonumber \\
&& \delta \tilde J_{a\mu} = \partial_\mu \alpha_a + \epsilon_{abc} \tilde J_{b\mu} \alpha_c \,
\label{su2.local}
\end{eqnarray}
Notice that in the above equation $\alpha_a$ is a local parameter.

In order to implement the classical local $\rm SU(2)_L$ invariance
at the quantum level, one needs to define the composite
operator $\phi_0$ in Equation (\ref{nl.constraint}) by coupling it  in the classical action
to an external source $K_0$ through
the term
\begin{eqnarray}
S_{ext} = \int d^Dx \, K_0 \phi_0 \,
\label{k0.src}
\end{eqnarray}
$K_0$ is invariant under $\delta$.


The important observation now is that the variation of
full one-particle irreducible (1-PI)
vertex functional $\G^{(0)} = S+S_{ext}$
is linear in the quantized fields $\phi_a$, \emph{i.e.},
\begin{eqnarray}
\delta \G^{(0)} = -\frac{1}{2} \int d^Dx \, \alpha_a(x)  K_0(x) \phi_a(x) \,
\label{local.var}
\end{eqnarray}
By taking a derivative of both sides of the above equation
w.r.t. $\alpha_a(x)$ one obtains the LFE for the tree-level
vertex functional $\G^{(0)}$:
\begin{align}
\!\!\!\!
{\cal W}_a(\G^{(0)}) & = 
-\partial_\mu \frac{\delta \G^{(0)}}{\delta \tilde J_{a\mu}} +
\epsilon_{acb} \tilde J_{c\mu} \frac{\delta \G^{(0)}}{\delta \tilde J_{b\mu}}+
\frac{1}{2} \frac{\delta \G^{(0)}}{\delta K_0(x)} \frac{\delta \G^{(0)}}{\delta \phi_a(x)} + \frac{1}{2} \epsilon_{abc} \phi_c(x) \frac{\delta \G^{(0)}}{\delta \phi_b(x)} 
\nonumber \\
& = -\frac{1}{2}  
K_0(x) \phi_a(x)
\label{gl.f.eq}
\end{align}
Notice that the $\phi_0$-term, entering in the variation of the $\phi_a$
field, is generated by $\frac{\delta \G^{(0)}}{\delta K_0(x)}$.
The advantage of this formulation resides in the fact that it is suitable
to be promoted at the quantum level. Indeed by defining the
composite operator $\phi_0$ by taking functional derivatives w.r.t.
its source $K_0$, one is able to control its renormalization, once
radiative corrections are included~\cite{ZinnJustin:1974mc}.

In the following Section we are going to give a compact and
self-contained presentation of the algebraic techniques used to deal with
bilinear functional equations like the LFE in Equation (\ref{gl.f.eq}).

\section{Ancestor Amplitudes and Weak Power-Counting}

We are going to discuss in this Section the consequences of the
LFE for the full vertex functional. The imposition of a quantum
symmetry in a non-power-counting renormalizable theory is a subtle
problem, since in general there is no control on the dimensions
of the possible breaking terms as strong as the one guaranteed by the
Quantum Action Principle (QAP) in the renormalizable case.
Let us discuss the latter case first.

\subsection{Renormalizable Theories and the Quantum Action Principle}

If the tree-level functional $\G^{(0)}$ is power-counting
renormalizable, the renormalization procedure~\cite{Velo:1976gh} provides
a way to compute all higher-order terms in the loop expansion of
the full vertex functional $\G[\Phi,\chi] = \sum_{n=0}^\infty \hbar^n \G^{(n)}[\Phi,\chi]$, depending on the set of quantized fields $\Phi$ and external sources
collectively denoted by $\chi$, by fixing order by order only a finite
set of action-like normalization conditions. One says that the classical action is therefore stable under radiative corrections, namely the number of free parameters does not increase with the loop order.

This procedure is a recursive one, since it allows to construct
$\G^{(n)}$ once $\G^{(j)}$, $j<n$ are known. From a combinatorial point of view, it turns out that $\G$ is the generating functional of the 1-PI renormalized
Feynman amplitudes.

A desirable feature of power-counting renormalizable theories is that
the dependence of 1-PI Green's functions under an infinitesimal
variations of the quantized fields and of the parameters of the model
is controlled by the so-called Quantum Action Principle 
(QAP)~\cite{Breitenlohner:1977hr}-\cite{Lowenstein:1971vf}
and can be
expressed as the insertion of certain {\em local} operators
with UV dimensions determined by their tree-level approximation
(\emph{i.e.}, a polynomial in the fields, the external sources and derivatives thereof).



Let us now consider a certain symmetry $\delta$ of the tree-level $\G^{(0)}$
classical action.
Under the condition that the symmetry $\delta$ is non-anomalous~\cite{Piguet:1995er}, it can be extended to the full vertex functional $\G$.
 In many cases of physical interest
the proof that the symmetry is non-anomalous can be performed by making use
of cohomological tools. Namely one writes the functional equation
associated with the $\delta$-invariance of the tree-level vertex functional
as follows:
\begin{eqnarray}
{\cal S}(\G^{(0)}) \equiv \int d^Dx \, \sum_{\Phi} \frac{\delta \G^{(0)}}{\delta \Phi(x)}
\frac{\delta \G^{(0)}}{\delta \Phi^*(x)} = 0
\label{cl.master.eq}
\end{eqnarray}
where $\Phi^*$ is an external source coupled in the tree-level
vertex functional to the $\delta$-transformation of $\Phi$
and the sum is over the quantized fields.
$\Phi^*$ are known as antifields~\cite{Gomis:1994he}.
If $\delta$ is nilpotent (as it happens, e.g., for the Becchi--Rouet--Stora--Tyutin (BRST) operator~\cite{Becchi:1975nq}-\cite{Becchi:1974xu} 
in gauge theories), the recursive proof of the absence of
obstructions to the fulfillment of Equation (\ref{cl.master.eq}) works as follows.
Suppose that Equation~(\ref{cl.master.eq}) is satisfied up to order $n-1$ in the loop
expansion. Then by the QAP the $n$-th order breaking
\begin{eqnarray}
\Delta^{(n)} & \equiv & \int d^Dx \, \sum_{\Phi}
\Big ( \frac{\delta \G^{(0)}}{\delta \Phi(x)}\frac{\delta \G^{(n)}}{\delta \Phi^*(x)}
+ \frac{\delta \G^{(n)}}{\delta \Phi(x)}\frac{\delta \G^{(0)}}{\delta \Phi^*(x)}
+ \sum_{j=1}^{n-1} \frac{\delta \G^{(j)}}{\delta \Phi(x)}
\frac{\delta \G^{(n-j)}}{\delta \Phi^*(x)} \Big ) \nonumber \\
\label{nthbrkg}
\end{eqnarray}
is a polynomial in the fields, the external sources and their derivatives.
The term involving $\G^{(n)}$ in Equation (\ref{nthbrkg}) allows to define
the linearized operator ${\cal S}_0$ according to
\begin{eqnarray}
{\cal S}_0(\G^{(n)}) \equiv \int d^Dx \, \sum_{\Phi}
\Big ( \frac{\delta \G^{(0)}}{\delta \Phi(x)}\frac{\delta \G^{(n)}}{\delta \Phi^*(x)}
+ \frac{\delta \G^{(n)}}{\delta \Phi(x)}\frac{\delta \G^{(0)}}{\delta \Phi^*(x)} \Big ) \,
\end{eqnarray}
${\cal S}_0$ is also nilpotent, as a consequence of the
nilpotency of $\delta$ and of the tree-level invariance in Equation~(\ref{cl.master.eq}).
By exploiting this fact and by  applying ${\cal S}_0$ on both sides of Equation (\ref{nthbrkg})
one finds
\begin{eqnarray}
{\cal S}_0 (\Delta^{(n)}) = 0 \,
\label{anomaly}
\end{eqnarray}
provided that the Wess-Zumino consistency condition~\cite{Wess:1971yu}
\begin{eqnarray}
{\cal S}_0  \Big ( \sum_{j=1}^{n-1} \frac{\delta \G^{(j)}}{\delta \Phi(x)}
\frac{\delta \G^{(n-j)}}{\delta \Phi^*(x)} \Big ) = 0
\label{wz}
\end{eqnarray}
holds. This is the case, e.g., for the BRST symmetry and the
associated master Equation (\ref{cl.master.eq}), since Equation (\ref{wz}) turns out to be
a consequence of a generalized Jacobi identity for the
Batalin-Vilkovisky bracket for the conjugated variables
$(\Phi,\Phi^*)$~\cite{Gomis:1994he}.

The problem of establishing whether the functional identity
\begin{eqnarray}
{\cal S}(\G)=0
\label{master.eq}
\end{eqnarray}
holds at order $n$ then boils down to prove that
the most general solution to Equation (\ref{anomaly}) is of the form
\begin{eqnarray}
\Delta^{(n)} = -{\cal S}_0(\Xi^{(n)})
\label{empty}
\end{eqnarray}
%


since then $\G^{'(n)} \equiv \G^{(n)} + \Xi^{(n)}$ will fulfill
Equation (\ref{master.eq}) at order $n$ in the loop expansion.
\emph{I.e.}, the problem reduces to the computation of the cohomology
$H({\cal S}_0)$ of the operator
${\cal S}_0$ in the space of integrated local polynomials in the fields,
the external sources and their derivatives.
Two ${\cal S}_0$-invariant
integrated local polynomials ${\cal J}_1$ and ${\cal J}_2$
belong to the same cohomology class in $H({\cal S}_0)$ if and only if
\begin{eqnarray}
{\cal J}_1 = {\cal J}_2 + {\cal S}_0({\cal K})
\label{cohom.expl1.}
\end{eqnarray}
for some integrated local polynomial ${\cal K}$. In particular,
$H({\cal S}_0)$ is empty if the only cohomology class
is the one of the zero element, so that the condition that
${\cal J}_1$ is ${\cal S}_0$-invariant implies that
\begin{eqnarray}
{\cal J}_1 = {\cal S}_0({\cal K})
\end{eqnarray}
for some ${\cal K}$.
Hence if one can prove that  the cohomology of the operator ${\cal S}_0$
is empty in the space
of breaking terms,
then Equation (\ref{empty}) must be fulfilled by some
choice of the functional $\Xi^{(n)}$.
Moreover it must be checked
that the UV dimensions of the possible counterterms $\Xi^{(n)}$ are compatible with the action-like condition, so that renormalizability of the theory is
not violated.
An extensive review of BRST cohomologies for gauge theories
is given in~\cite{Barnich:2000zw}.

\subsection{Non-Renormalizable Theories}

The QAP does not in general hold for non-renormalizable theories.
This does not come as a surprise, since the appearance of UV
divergences with higher and higher degree, as one goes up with the loop
order, prevents to characterize the induced breaking of a functional identity in terms of a polynomial of a given finite degree (independent of the loop order).

Moreover for the NLSM another important difference must be stressed: the basic Green's functions of the theory are not those of the quantized fields $\phi_a$, but those of the flat connection coupled to the external vector source
$\tilde J_{a\mu}$ and of the non-linear constraint $\phi_0$ (coupled to $K_0$).
This result follows from the invertibility of
$$\frac{\delta \G}{\delta K_0}=\phi_0 + O(\hbar)$$
as a formal power series in $\hbar$ (since
$\left . \phi_0 \right |_{\phi_a=0} = m_D$).
Then the LFE for the vertex functional $\G$
\begin{eqnarray}
{\cal W}_a(\G) = -\frac{1}{2} K_0(x) \phi_a(x)
\label{lfe}
\end{eqnarray}
can be seen as a first-order functional differential equation
controlling the dependence of $\G$ on the fields $\phi_a$.
Provided that a solution exists (as will be proven in Section~\ref{sec:cohom}),
Equation (\ref{lfe}) determines all the amplitudes involving at least
one external $\phi_a$-leg in terms of the boundary condition
provided by the functional $\G[\tilde J,K_0] = \left .
\G[\phi,\tilde J, K_0] \right |_{\phi_a=0}$.

$\G[\tilde J,K_0]$ is the generating functional of
the so called ancestor amplitudes, \emph{i.e.}, the 1-PI amplitudes
involving only external $\tilde J$ and $K_0$ legs.


It is therefore reasonable to assume the LFE in Equation (\ref{lfe}) as
the starting point for the quantization of the theory.

From a path-integral point of view, Equation (\ref{lfe}) implies that
one is performing an integration over the SU(2)-invariant
Haar measure of the group, namely one is computing
\begin{eqnarray}
Z[J, \tilde J_\mu, K_0] & = &
\int {\cal D} \Omega(\phi) \exp \Big ( i \G^{(0)}[\phi,\tilde J_\mu,K_0] +
i \int d^Dx \, J_a \phi_a \Big ) \,
\label{path.integral.q}
\end{eqnarray}
where we denote by ${\cal D} \Omega(\phi)$ the SU(2) Haar measure
(in the coordinate representation spanned by the fields $\phi_a$).
This clarifies the geometrical meaning of the LFE.

\subsection{Weak Power-Counting}


As we have already noticed, in four dimensions
the NLSM is non power-counting renormalizable,
since already at one loop level an infinite number of
divergent $\phi$-amplitudes exists. One may wonder whether
the UV behavior of the ancestor amplitudes (the boundary
conditions to the LFE) is better.
It turns out that this is indeed the case and one finds that in $D$ dimensions
a $n$-th loop Feynman amplitude ${\cal G}$ with $N_{K_0}$ external $K_0$-legs and
$N_{\tilde J}$ external $\tilde J$-legs has superficial degree
of divergence given by~\cite{Ferrari:2005va}
\begin{eqnarray}
d({\cal G}) \leq (D-2) n + 2 - N_{\tilde J} - 2 N_{K_0} \,
\label{wpc}
\end{eqnarray}
The proof is straightforward although somehow lengthy and will not
be reported here. It can be found in~\cite{Ferrari:2005va}.
Equation~(\ref{wpc}) establishes the Weak Power-Counting (WPC) condition:
at every loop order only a finite number of superficially divergent ancestor
amplitudes exist.

For instance, in $D=4$ and at one loop order, Equation~(\ref{wpc}) reduces
to
\begin{eqnarray}
d ({\cal G}) \leq 4 -  N_{\tilde J} - 2 N_{K_0} \,
\label{wpc.1.loop}
\end{eqnarray}
\emph{i.e.}, UV divergent amplitudes involve only up to four external $\tilde J_\mu$ legs or two $K_0$-legs.


By taking into account Lorentz-invariance and global $\rm SU(2)_R$ symmetry, the list of UV divergent amplitudes reduces to
\begin{align}
& 
\!\!\!
\int d^4x \, \partial_\mu \tilde J_{a\nu} \partial^\mu \tilde J_a^\nu \, ,
~~
\int d^4x \, (\partial\tilde J_a)^2 \, ,
~~
\int d^4x \, \epsilon_{abc} \partial_\mu \tilde J_{a\nu}
\tilde J^\mu_b \tilde J^\nu_c \, ,
~~
\int d^4x \, (\tilde J_a)^2 (\tilde J_b)^2 \, 
\nonumber \\
&
\int d^4x \, {\tilde J}_{a\mu} {\tilde J}_b^\mu {\tilde J}_{a\nu} {\tilde J}_b^\nu \, ,
~~
\int d^4x \, {\tilde J}_{a\mu}^2 \, , ~~
\int d^4x \, K_0^2 \, , ~~
\int d^4x \, K_0 \tilde J_a^2 \,
\label{1.loop.div}
\end{align}
Notice that the counterterms are local.

It should be emphasized that the model is not power-counting renormalizable,
even when ancestor amplitudes are considered, since according to Equation (\ref{wpc}) the number of UV divergent amplitudes increases as the loop
order $n$ grows.

\medskip
A special case is the 2-dimensional NLSM. For $D=2$ Equation (\ref{wpc})
yields
\begin{eqnarray}
d({\cal G}) \leq 2 - N_{\tilde J} - 2 N_{K_0}
\end{eqnarray}
%

\noindent \emph{i.e.}, at every loop order there can be only two UV divergent
ancestor amplitudes, namely
$$\int d^2x \, \tilde J^2 \qquad \mbox{and}   \qquad
  \int d^2x \, K_0$$
These are precisely of the same functional form as the ancestor
amplitudes entering in the tree-level vertex functional and,
in this sense, the model shares the stability property of the classical
action typical of power-counting renormalizable models.
Renormalizability of the 2-dimensional NLSM
can also be established by relying on the  Ward identity
of global SU(2) symmetry (see e.g.,~\cite{ZinnJustin:2002ru}).

A comment is in order here. In References~\cite{Weinberg:1978kz,Gasser:1983yg}
the external fields are the sources of connected Green's functions
of certain quark-antiquark currents. The ancestor amplitudes
in the NLSM, in the approach based on the LFE, do not have a direct physical
interpretation of this type, however they have a very clear geometrical
meaning.
First of all, $\tilde{J}_\mu$ is the source coupled to the flat
connection naturally associated with the group element $\Omega$.
On the other hand, $K_0$ is the unique scalar source required,
in the special case of the SU(2) group, in order to control
the renormalization of the non-linear classical SU(2)
transformation of the $\phi_a$'s and thus plays the role of the
so-called antifields~\cite{Gomis:1994he,ZinnJustin:1974mc}.
The extension to a general Lie group $G$ is addressed at the
end of Section~\ref{sec:cohom}.

\section{Cohomological Analysis of the LFE}\label{sec:cohom}


In order to study the properties of the LFE, it is very convenient
to introduce a fictious BRST operator $s$ by promoting the gauge
parameters $\alpha_a(x)$ to classical anticommuting ghosts
$\omega_a(x)$.
\emph{I.e.}, one sets
\begin{align}
& s \tilde J_{a\mu} = \partial_\mu \omega_a + \epsilon_{abc}
\tilde J_{b\mu} \omega_c \, , \quad
s \phi_a = \frac{1}{2} \omega_a \phi_0 + \frac{1}{2} \epsilon_{abc} \phi_b \omega_c \, , \qquad
s \phi_0 = -\frac{1}{2} \omega_a \phi_a \,
\nonumber \\
&
s K_0 = \frac{1}{2} \omega_a \frac{\delta \G^{(0)}}{\delta \phi_a(x)} \, ,
\quad
 s \omega_a = -\frac{1}{2} \epsilon_{abc} \omega_b \omega_c \,
\label{brst}
\end{align}
Some comments are in order here. First of all the BRST operator
$s$ acts also on the external source $K_0$.
Moreover, the BRST transformation of $\omega_a$ is fixed by nilpotency,
namely $s^2 = 0$.

The introduction of the ghosts allows to define a grading w.r.t. the conserved
ghost
number. $\omega$ has ghost number $+1$, while all the other
fields and sources have ghost number zero. (The ghost number was called the Faddeev-Popov ($\Phi\Pi$) charge in~\cite{Ferrari:2005va}.)

In terms of the operator $s$ we can write the $n$-th order projection ($n \geq 1$)
of the LFE in Equation (\ref{lfe}) as follows:
\begin{eqnarray}
\Big [ \int d^Dx \, \omega_a {\cal W}_a(\G) \Big ]^{(n)} =
s \G^{(n)} + \sum_{j=1}^{n-1} \int d^Dx \, \frac{1}{2} \omega_a
\frac{\delta \G^{(j)}}{\delta K_0} \frac{\delta \G^{(n-j)}}{\delta \phi_a} = 0
\label{lfe.nth.order}
\end{eqnarray}
Notice that the bilinear term in the LFE manifests itself into
the presence of the mixed $\frac{\delta \G^{(j)}}{\delta K_0} \frac{\delta \G^{(n-j)}}{\delta \phi_a}$ contribution. Moreover in the r.h.s. there is no contribution from the breaking term linear in $\phi_a$ in Equation (\ref{gl.f.eq}) since the latter remains classical.

Suppose now that all divergences have been recursively subtracted
up to order $n-1$. At the $n$-th order the UV divergent part
can only come from the term involving $\G^{(n)}$ in Equation (\ref{lfe.nth.order})
and therefore, if the LFE holds, one gets a condition on the UV divergent
part $\G^{(n)}_{pol}$ of $\G^{(n)}$:
\begin{eqnarray}
s \G^{(n)}_{pol} = 0 \,
\label{pol}
\end{eqnarray}
To be specific, one can use Dimensional Regularization and subtract
only the pole part of the ancestor amplitudes (after the proper normalization
of the ancestor background connection amplitudes
$$\frac{m}{m_D} \frac{\delta^{(n)}\G}{\delta {\tilde J}^{\mu_1}_{a_1} \dots
\delta {\tilde J}^{\mu_n}_{a_n}} \, $$
The LFE then fixes the correct factor for the normalization of amplitudes
involving $K_0$).
This subtraction procedure has been shown
to be symmetric~\cite{Ferrari:2005va,Bettinelli:2007zn}, \emph{i.e.}, to preserve the
LFE. The pole parts before subtraction obey the condition
in Equation (\ref{pol}).


By the nilpotency of $s$, solving Equation (\ref{pol}) is equivalent
to computing the cohomology of the BRST operator $s$ in the
space of local functionals in $\tilde J, \phi, K_0$ and
their derivatives with ghost number zero.
This can be achieved by using the techniques developed
in~\cite{Henneaux:1998hq}.

One first builds invariant combinations in one-to-one correspondence
with the ancestor variables $\tilde J_{a\mu}$ and $K_0$.
For that purpose it is more convenient to switch back to matrix notation.
The difference \linebreak $I_\mu \equiv F_\mu - \tilde J_\mu$ transforms in the adjoint representation of SU(2), being the difference of two gauge connections. Thus the conjugate
of such a difference w.r.t. $\Omega$
\begin{eqnarray}
j_\mu = j_{a\mu} \frac{\tau_a}{2} = \Omega^\dagger I_\mu \Omega
\label{j}
\end{eqnarray}
is invariant under $s$. By direct computation one finds
\begin{eqnarray}
m_D^2 j_{a\mu} & = & m_D^2 I_{a\mu} - 2 \phi_b^2 I_{a\mu} +
2 \phi_b I_{b\mu} \phi_a + 2 \phi_0 \epsilon_{abc} \phi_b I_{c\mu}
\nonumber \\
& \equiv & m_D^2 R_{ba} I_{b\mu}
\label{j.comp}
\end{eqnarray}
The matrix $R_{ba}$ is an element of the adjoint representation
of SU(2) and therefore the mapping $\tilde J_{a\mu} \rightarrow
j_{a\mu}$ is invertible.

One can also prove that the following combination
\begin{eqnarray}
\overline{K}_0 \equiv
\frac{m_D^2 K_0}{\phi_0} - \phi_a \frac{\delta S}{\delta \phi_a}
\label{k0}
\end{eqnarray}
is invariant~\cite{Ferrari:2005va}.
At $\phi_a=0$ one gets
\begin{eqnarray}
\left . \overline{K}_0 \right |_{\phi_a=0} = m_D K_0 \,
\label{k0.ic.}
\end{eqnarray}
and therefore the transformation
$K_0 \rightarrow \overline{K}_0$  is also invertible.

In terms of the new variables $\overline{K}_0$ and $j_\mu$
and by differentiating Equation (\ref{pol}) w.r.t. $\omega_a$
one gets
\begin{eqnarray}
\Theta_{ab} \frac{\delta \G^{(n)}_{pol}[j,\overline K, \phi]}{\delta \phi_b} = 0
\label{pol.bleached}
\end{eqnarray}
%

\noindent where $s \phi_b = \omega_ a\Theta_{ab} $, \emph{i.e.},
\begin{eqnarray}
\Theta_{ab} = \frac{1}{2} \phi_0 \delta_{ab} + \frac{1}{2} \epsilon_{abc} \phi_c \,
\label{brst.phi}
\end{eqnarray}
$\Theta_{ab}$ is invertible and thus Equation (\ref{pol.bleached}) yields
\begin{eqnarray}
\frac{\delta \G^{(n)}_{pol}[j,\overline K_0, \phi]}{\delta \phi_b} = 0 \,
\end{eqnarray}
This equation is a very powerful one. It states that the $n$-th order
divergences (after the theory has been made finite up to order $n-1$)
of the $\phi$-fields can only appear through the invariant
combinations $\overline{K}_0$ and $j_{a\mu}$. These invariant
variables have been called bleached variables and they are in one-to-one correspondence with the ancestor variables $K_0$ and $\tilde J_{a\mu}$.

The subtraction strategy is thus the following. One computes the
divergent part of the properly normalized ancestor amplitudes
that are superficially divergent at a given loop order
according to the WPC formula in Equation (\ref{wpc}).
Then the replacement $\tilde J_{a\mu} \rightarrow j_{a\mu}$ and
$K_0 \rightarrow \overline{K}_0$ is carried out. This gives
the full set of counterterms required to make the theory finite
at order $n$ in the loop expansion.

As an example, we give here the explicit form of the
one-loop divergent counterterms for the NLSM in $D=4$
~\cite{Ferrari:2005va} (notice that we have set $g=1$
according to our conventions in this paper):
\begin{eqnarray}
\hat \G^{(1)} & = & \frac{1}{D-4} \Big [
- \frac{1}{12} \frac{1}{(4\pi)^2} \frac{m_D^2}{m^2} \Big (
{\cal I}_1 - {\cal I}_2 -  {\cal I}_3 \Big )
+  \frac{1}{(4\pi)^2} \frac{1}{48} \frac{m_D^2}{m^2}
\Big ( {\cal I}_6 + 2 {\cal I}_7 \Big ) \nonumber \\
& & +
\frac{1}{(4\pi)^2} \frac{3}{2} \frac{1}{m^2 m_D^2} {\cal I}_4
+ \frac{1}{(4\pi)^2} \frac{1}{2} \frac{1}{m^2} {\cal I}_5 \Big ] \,
\label{oneloop.cts}
\end{eqnarray}
By projecting the above equation on the relevant monomial in the $\phi_a$ fields
one can get the divergences of the descendant amplitudes.
As an example, for the four point $\phi_a$ function one gets
by explicit computation that the contribution from the combination
${\cal I}_1 - {\cal I}_2 - {\cal I}_3$ is zero, while the remaining
invariants give
\begin{eqnarray}
\!\!\!\!\!\!\!
\hat \G^{(1)} [\phi\phi\phi\phi] & = & - \frac{1}{D-4}
\frac{1}{m_D^2 m^2 (4 \pi)^2} \nonumber \\
&& ~~~ \int d^D x \,
\Big (
- \frac{1}{3} \partial_\mu \phi_a \partial^\mu \phi_a
\partial_\nu \phi_b \partial^\nu \phi_b
- \frac{2}{3} \partial_\mu \phi_a \partial_\nu \phi_a
              \partial^\mu \phi_b \partial^\nu \phi_b
\nonumber \\
&& ~~~~
-\frac{3}{2} \phi_a \square \phi_a \phi_b \square \phi_b - 2 \phi_a \square \phi_a \partial_\mu \phi_b \partial^\mu \phi_b
\Big ) \, .
\label{4phi}
\end{eqnarray}

The invariants in the combination ${\cal I}_6 + 2 {\cal I}_7$
generate the counterterms in the first line between square
brackets; these counterterms are globally SU(2) invariant. The other terms
are generated by invariants involving the source $K_0$.
In \cite{Appelquist:1980ae,Tataru:1975ys} they were
constructed  by means of a (non-locally invertible)
field redefinition of $\phi_a$.
The full set of mixed four point amplitudes involving at least
one $\phi_a$ legs and the external sources $\tilde J_\mu$ and
$K_0$ can be found in~\cite{Ferrari:2005va}.

The correspondence with the linear sigma model in the large
coupling limit has been studied in~\cite{Bettinelli:2006ps}.

The massive NLSM in the LFE formulation has been studied in~\cite{Ferrari:2010ge}, while the symmetric subtraction procedure for the LFE associated with polar coordinates in the simplest case of the free complex scalar field has been given in~\cite{Ferrari:2009uj}.

In the SU(2) NLSM just one scalar source $K_0$ is
sufficient in order to formulate the LFE. For an arbitrary
Lie group $G$ the LFE can always be written if one introduces
a full set of antifields $\phi^*_I$, as follows.
Let us denote by $\Omega(\phi_I)$ the group element
belonging to $G$, parameterized by local coordinates
$\phi_I$.
Then under an infinitesimal left $G$-transformation
of parameters $\alpha_J$
\begin{eqnarray}
\delta \Omega = i \alpha_J T_J \Omega \,
\end{eqnarray}
where $T_J$ are the generators of the group $G$, one has
\begin{eqnarray}
\delta \phi_I = S_{IJ}(\phi) \alpha_J \,
\end{eqnarray}
It is convenient to promote the local left invariance to a BRST
symmetry by upgrading the parameters $\alpha_J$ to local
classical anticommuting ghosts $C_J$. Then one can
introduce in the usual way the couplings with the
antifields $\phi_I^*$ through
\begin{eqnarray}
S_{ext} = \int d^Dx \, \phi^*_I S_{IJ}(\phi) C_J \,
\end{eqnarray}
and then write the corresponding BV master equation~\cite{Gomis:1994he}.
This is the generalization of the
LFE valid for the group $G$. The cohomology of the linearized
BV operator (which is the main tool for identifying the
bleached variables, as shown above) has been studied
for any Lie group $G$ in~\cite{Henneaux:1998hq}.

\section{Higher Loops}

At orders $n>1$ the LFE for $\G^{(n)}$ is an inhomogeneous equation
\begin{eqnarray}
s \G^{(n)} = \Delta^{(n)} \equiv -\frac{1}{2} \int d^Dx \, \omega_a \sum_{j=1}^{n-1} \frac{\delta \G^{(j)}}{\delta K_0}
\frac{\delta \G^{(n-j)}}{\delta \phi_a} \,
\label{hl.1}
\end{eqnarray}
The above equation can be explicitly integrated by using the techniques
of the Slavnov-Taylor (ST) parameterization of the effective 
action~\cite{Quadri:2005pv}-\cite{Quadri:2003ui}
(originally developed in order to provide a strategy for the restoration
of the ST identity of non-anomalous gauge theories in the absence of a symmetric
regularization).

For that purpose it is convenient to redefine the ghost according to
\begin{eqnarray}
\overline{\omega}_a = \Theta_{ab} \omega_b
\label{hl.ghost}
\end{eqnarray}
where $\Theta_{ab}$ is given in Equation (\ref{brst.phi}). The action of $s$ then reduces to
\begin{eqnarray}
s \overline{K}_0 = s j_{a\mu} = 0 \, , \qquad s \phi_a = \overline{\omega}_a \, ,
\qquad s \overline{\omega}_a = 0 \,
\label{hl.brst.triv}
\end{eqnarray}
This means that the variables $\overline{K}_0$ and $j_{a\mu}$ are invariant,
while the pair $(\phi_a, \overline{\omega}_a)$ is a BRST doublet (\emph{i.e.},
a pair of variables $u,v$ such that $s~u = v, s~v =0$)~\cite{Gomis:1994he,Quadri:2002nh}.


By the nilpotency of $s$ the following consistency condition must hold for $\Delta^{(n)}$:
\begin{eqnarray}
s \Delta^{(n)} =0 \,
\label{hl.cons.cond}
\end{eqnarray}
The fulfillment of the above equation as a consequence of the validity of the LFE up to
order $n-1$ is proven in~\cite{Quadri:2005pv}.
In terms of the new variables Equation (\ref{hl.1}) reads
\begin{eqnarray}
\int d^Dx \, \overline{\omega}_a \frac{\delta \G^{(n)}}{\delta \phi_a} =
\Delta^{(n)}[\overline{\omega}_a,\phi_a,\overline{K}_0,j_{a\mu}] \,
\label{hl.new.var}
\end{eqnarray}
By noticing that $\Delta^{(n)}$ is linear in $\overline{\omega}_a$
and by differentiating Equation (\ref{hl.new.var}) w.r.t.
$\overline{\omega}_a$ we arrive at
\begin{eqnarray}
\frac{\delta \G^{(n)}}{\delta \phi_a(x)} = \frac{\delta \Delta^{(n)}}{\delta \overline{\omega}_a(x)} \,
\label{hl.diff.eq}
\end{eqnarray}
The above equation controls the explicit dependence of the $n$-th order
vertex functional on $\phi_a$ (there is in addition an implicit dependence
on $\phi_a$ through the variables $j_{a\mu}$ and $\overline{K}_0$).

The explicit dependence on $\phi_a$ only appears through lower order terms.
Hence it does not influence the $n$-th order ancestor amplitudes.

The solution of Equation (\ref{hl.1}) can be written in compact form
by using a homotopy operator. Indeed $\G^{(n)}$ will be the
sum of a $n$-th order contribution ${\cal A}^{(n)}$, depending only on
$j_{a\mu}$ and $\overline{K}_0$, plus a lower order~term:
\begin{eqnarray}
\G^{(n)}[\phi_a,K_0,\tilde J_{a\mu}] & = &
{\cal A}^{(n)}[\overline{K_0},j_{a\mu}] \nonumber \\
& & + \int d^Dx \int_0^1 dt ~ \phi_a(x) \lambda_t
\frac{\delta \Delta^{(n)}}{\delta \overline{\omega}_a(x)} \,
\label{hl.exp,sol}
\end{eqnarray}
The operator $\lambda_t$ acts as follows on a generic functional
$X[\phi_a, \overline{\omega}_a, \overline{K}_0, j_{a\mu}]$:
\begin{eqnarray}
\lambda_t X [\phi_a, \overline{\omega}_a, \overline{K}_0, j_{a\mu}] =
X[t \phi_a, t \overline{\omega}_a,\overline{K}_0, j_{a\mu}] \,
\label{hl.lambda}
\end{eqnarray}
The homotopy operator $\kappa$ for the BRST differential $s$
in the second line of Equation (\ref{hl.exp,sol})
is therefore given by
\begin{eqnarray}
\kappa = \int d^Dx \, \int_0^1 dt \,\phi_a(x) \lambda_t \frac{\delta}{\delta \overline{\omega}_a(x)}
\label{hl.hom}
\end{eqnarray}
and satisfies the condition
\begin{eqnarray}
\{ s, \kappa \} = {\bf 1}
\label{hl.comm}
\end{eqnarray}
where ${\bf 1}$ denotes the identity on the space of functionals
spanned by $\overline{\omega}_a, \phi_a$.

\medskip
An important remark is in order here. The theory remains finite
and respects the LFE if one adds to $\G^{(n)}$
some
integrated local monomials  in $j_{a\mu}$
 and $\overline{K}_0$ and ordinary derivatives thereof (with finite coefficients), compatible
with Lorentz symmetry and global SU(2) invariance,
while respecting the WPC condition in Equation (\ref{wpc}):
\begin{eqnarray}
\G^{(n)}_{finite} = \sum_j \int d^Dx \,  {\cal M}_j(j_{a\mu}, \overline{K}_0) \,
\label{finite.ren}
\end{eqnarray}
This is a consequence of the non power-counting renormalizability of the theory:
one can introduce order by order in the loop expansion an increasing number
of finite parameters that do not appear in the classical action.
Notice that they cannot be inserted back at tree-level: if one performs such
an operation, the WPC condition is lost.

This observation suggests that these finite parameters cannot be easily understood
as physical free parameters of the theory, since they cannot appear
in the tree-level action. It was then proposed to define the model
by choosing the symmetric subtraction scheme discussed in Section~\ref{sec:cohom} and
by considering as physical parameters only those present in the classical
action plus the scale of the radiative corrections $\Lambda$~\cite{Bettinelli:2007zn}.
While acceptable on physical grounds, from the mathematical point of view
one may wonder whether there is some deeper reason justifying such a strategy.
We will comment briefly on this point in the Conclusions.

\section{Applications to Yang-Mills and the Electroweak Theory}

When the vector source $\tilde J_{a\mu}$ becomes
a dynamical gauge field, the NLSM action
gives rise to the St\"uckelberg mass term~\cite{Ferrari:2004pd}.

The subtraction procedure based on the LFE has been
used to implement a mathematically consistent formulation
of non-linearly realized massive Yang-Mills theory.
SU(2) Yang-Mills in the LFE formalism
has been formulated in~\cite{Bettinelli:2007tq}.
The pseudo-Goldstone fields take over the role of the $\phi_a$
fields of the NLSM. Their Green's functions are fixed
by the LFE. The WPC proves to be very restrictive, since
by imposing the WPC condition it turns out that the only
allowed classical solution is the usual Yang-Mills theory plus the
St\"uckelberg mass term.

This is a very powerful (and somehow surprising) result. Indeed
all possible monomials constructed out of $j_{a\mu}$ and
ordinary derivatives thereof are gauge-invariant and
therefore they could be used as interaction vertices in the
classical action.

Otherwise said, the peculiar structure of the Yang-Mills action
\begin{eqnarray}
S_{\rm YM} = - \int d^4x \, \frac{1}{4} G_{a\mu\nu}G_a^{\mu\nu} \,
\label{ym}
\end{eqnarray}
where $G_{a\mu\nu}$ denotes the field strength of the gauge field $A_{a\mu}$
$$G_{a\mu\nu} = \partial_\mu A_{a\nu} - \partial_\nu A_{a\mu} + f_{abc} A_{b\mu} A_{c\nu}$$
is not automatically enforced by the requirement of gauge invariance if
the gauge group is non-linearly realized.
However if the WPC condition is satisfied, the only admissible solution
becomes Yang-Mills theory plus the St\"uckelberg mass term:
\begin{eqnarray}
S_{\rm nlYM} = S_{\rm YM} + \int d^4x \, \frac{M^2}{2} (A_{a\mu} - F_{a\mu})^2 \,
\label{stueck}
\end{eqnarray}
Massive Yang-Mills theory in the presence of a non-linearly
realized gauge group is physically unitary~\cite{Ferrari:2004pd}
(despite the fact that it violates the Froissart bound~\cite{Froissart:1961ux}
-\cite{Bettinelli:2011rx}
at tree-level).
The counterterms in the Landau gauge  have been computed at one loop level in~\cite{Bettinelli:2007cy}.
The formulation of the theory in a general 't Hooft gauge has been
given in~\cite{Bettinelli:2007eu}.

The approach based on the LFE can also be used for non-perturbative
studies of Yang-Mills theory on the lattice.
The phase diagram of SU(2) Yang-Mills has been considered in~\cite{Ferrari:2011aa}.
Emerging evidence is being accumulated about the formation of
isospin scalar bound states~\cite{Ferrari:2013lja} in the supposedly confined phase of the theory~\cite{Ferrari:2011bx}.

An analytic approach based on the massless bound-state formalism
for the implementation of the Schwinger mechanism in non-Abelian gauge theories
has been presented in~\cite{Aguilar:2011xe}-\cite{Ibanez:2012zk}.

A very important physical application of non-linearly realized gauge theories
is the formulation of a non-linearly realized electroweak theory, based
on the group $\rm SU(2) \times U(1)$.
The set of gauge fields comprises the  SU(2) fields $A_{a\mu}$
and the hypercharge U(1) gauge connection $B_\mu$.
By using the technique of bleached variables one can first
construct SU(2) invariant variables in one-to-one correspondence
with $A_\mu = A_{a\mu} \frac{\tau_a}{2}$~\cite{Bettinelli:2007eu}:
\begin{eqnarray}
w_\mu =  \Omega^\dagger g A_\mu \Omega - g' \frac{\tau_3}{2} B_\mu
+ i \Omega ^\dagger \partial_\mu \Omega \equiv w_{a\mu} \frac{\tau_a}{2} \,
\label{w}
\end{eqnarray}
In the above equation we have reinserted back for later convenience
the SU(2) and U(1) coupling
constants $g$ and $g'$.
Since $w_\mu$ is SU(2) invariant, the hypercharge generator coincides with
the electric charge generator.
$w_{3\mu}$ is then the bleached counterpart of the $Z_\mu$ field,
since
\begin{eqnarray}
Z_\mu = \left . \frac{1}{\sqrt{g^2 + {g'}^2}} w_{3\mu} \right |_{\phi_a=0}
= c_W A_{3\mu} - s_W B_\mu
\label{z}
\end{eqnarray}
where $s_W$ and $c_W$ are the sine and cosine of the Weinberg angle
\begin{eqnarray}
s_W = \frac{g'}{\sqrt{g^2 + {g'}^2}} \, , \qquad
c_W = \frac{g}{\sqrt{g^2 + {g'}^2}} \,
\label{sw}
\end{eqnarray}
The photon $A_\mu$ is described by the combination orthogonal to $Z_\mu$,
namely
\begin{eqnarray}
A_\mu = s_W A_{3\mu} + c_W B_\mu \,
\label{photon}
\end{eqnarray}
One can built out of $A_{1\mu}$ and $A_{2\mu}$ the charged
$W^\pm$ field
\begin{eqnarray}
W^\pm_\mu = \frac{1}{\sqrt{2}} (A_{1\mu} \mp i A_{2\mu})
\label{W}
\end{eqnarray}
whose bleached counterpart is simply
\begin{eqnarray}
w^\pm_\mu = \frac{1}{\sqrt{2}} (w_{1\mu} \mp i w_{2\mu}) \,
\label{W.bleached}
\end{eqnarray}
The WPC allows for the same symmetric couplings of the
Standard Model and for two independent
mass invariants~\cite{Bettinelli:2008ey}-\cite{Bettinelli:2009wu}
%
\begin{eqnarray}
M_W^2 w^+w^- + \frac{M_Z^2}{2} w_{3\mu}^2
\label{2mass.inv}
\end{eqnarray}
where the mass of the $Z$ and $W$ bosons are not related by the
Weinberg relation
$$M_Z = \frac{M_W}{c_W} \, $$
This is a peculiar signature of the mass generation mechanism
{\em \`a la} St\"uckelberg, that is not present in
the linearly realized theory {\em \`a la} 
Brout-Englert-Higgs~\cite{Higgs:1964ia}-\cite{Englert:1964et}
(even if one
discards the condition of power-counting renormalizability
in favour of the WPC)~\cite{Quadri:2010uk}.

The inclusion of physical scalar resonances in the
non-linearly realized electroweak model, while respecting the
WPC, yields some definite prediction for the Beyond the Standard
Model (BSM) sector. Indeed it turns out that it is impossible
to add a scalar singlet without breaking the WPC condition.
The minimal solution requires
a SU(2) doublet of scalars, leading to a CP-even physical field
(to be identified with the recently
discovered scalar resonance at $125.6~GeV$) and to three additional
heavier physical states, one CP-odd and neutral and two charged
ones~\cite{Binosi:2012cz}.
The proof of the WPC in this model and the BRST identification
of physical states has been given in~\cite{Bettinelli:2013hia}.

The WPC and the symmetries of the theory select uniquely the tree-level
action of the non-linearly realized electroweak model.
As in the NLSM case, mathematically additional finite counterterms
are allowed at higher orders in the loop expansion.
In~\cite{Bettinelli:2007zn} it has been argued that they cannot be interpreted
as additional physical parameters (unlike in the effective
field theory approach), on the basis of the observation
that they are forbidden at tree-level by the WPC, and
this strategy has been consistently applied
in~\cite{Bettinelli:2007cy,Bettinelli:2009wu}.

The question remains open of whether a Renormalization Group
equation exists, involving a finite change in the higher
order subtractions, in such a way to compensate the change in
the sliding scale $\Lambda$ of the radiative corrections.
We notice that in this case the finite higher order counterterms
would be a function of the tree-level parameters only (unlike
in the conventional effective field theory approach, where they are
treated as independent extra free parameters).
This issue deserves further investigation, since obviously the
possibility of running the scale $\Lambda$ in a mathematically
consistent way would allow to obtain physical predictions
of the same observables applicable in different energy regimes.


\section{Conclusions}

The LFE makes it apparent that the independent amplitudes
of the NLSM are not those of the quantum fields, over which
the path-integral is carried out, but rather those of the
background connection $\tilde J_\mu$ and of the source $K_0$, coupled
to the solution of the non-linear constraint $\phi_0$.
The WPC can be formulated only for these ancestor amplitudes;
the LFE in turn fixes the descendant amplitudes, involving
at least one pion external leg. Within this formulation,
the minimal symmetric subtraction discussed in Section~\ref{sec:cohom}
is natural, since it provides a way to implement the idea
that the number of ancestor interaction vertices, appearing in the
classical action and compatible with the WPC, must be finite.


However, it should be stressed that the most general solution
to the LFE, compatible with the WPC, does not forbid to choose
different finite parts of the higher order symmetric
counterterms (as in the most standard view of effective field
theories, where such arbitrariness is associated with
extra free parameters of the non-renormalizable theory),
as far as they are introduced at the order prescribed by the WPC
condition and without violating the LFE.


In this connection it should be noticed that the addition of
 the symmetric finite renormalizations in Equation (\ref{finite.ren}),
that are allowed by the symmetries of the theory, is equivalent
to a change in the Hopf algebra~\cite{Connes:1999yr,Connes:2000fe} of the model.
This is because the finite counterterms in Equation (\ref{finite.ren})
modify the set of 1-PI Feynman diagrams on which the
Hopf algebra is constructed, as a dual of the enveloping algebra
of the Lie algebra of Feynman graphs.
The approach to renormalization based on Hopf algebras
is known to be  equivalent~\cite{EbrahimiFard:2010yy} to the traditional approach based on the
Bogoliubov recursive formula and its explicit solution through
the Zimmermann's forest formula~\cite{Zimmermann:1969jj}. For models endowed with a WPC
it might provide new insights into the structure of the UV
divergences of the theory. This connection seems to deserve
further investigations.


\section*{Acknowledgements}

It is a pleasure to acknowledge many enlightening discussions
with R.~Ferrari.
Useful comments and a careful reading of the manuscript by D.~Bettinelli are
also gratefully
acknowledged.



\pagebreak

\section*{\noindent Appendix: One-loop invariants}\label{app:A}
\vspace {12pt}

We report here the invariants controlling the one-loop divergences
of the NLSM in $D=4$ \cite{Ferrari:2005va}.
\begin{eqnarray}
&& {\cal I}_1 = \int d^Dx \, \Big [ D_\mu ( F -\tilde J )_\nu \Big ]_a \Big [ D^\mu ( F -\tilde J )^\nu \Big ]_a  \, ,
\nonumber \\
&& {\cal I}_2 = \int d^Dx \, \Big [ D_\mu ( F -\tilde J )^\mu \Big ]_a \Big [ D_\nu ( F -\tilde J )^\nu \Big ]_a  \, ,
\nonumber \\
&& {\cal I}_3 = \int d^Dx \, \epsilon_{abc} \Big [ D_\mu ( F - \tilde J )_\nu \Big ]_a \Big ( F^\mu_b - \tilde J^\mu_b \Big ) \Big ( F^\nu_c - \tilde J^\nu_c \Big ) \, ,  \nonumber \\
&& {\cal I}_4 = \int d^Dx \, \Big ( \frac{m_D^2 K_0}{\phi_0} - \phi_a \frac{\delta S}{\delta \phi_a} \Big )^2 \, , \nonumber \\
&& {\cal I}_5 = \int d^Dx \, \Big ( \frac{m_D^2 K_0}{\phi_0} - \phi_a \frac{\delta S}{\delta \phi_a} \Big ) \Big ( F^\mu_b - \tilde J^\mu_b \Big )^2 \, ,
\nonumber \\
&& {\cal I}_6 = \int d^Dx \, \Big ( F^\mu_a - \tilde J^\mu_a\Big  )^2
 \Big ( F^\nu_b - \tilde J^\nu_b \Big )^2 \, , \nonumber \\
&& {\cal I}_7 = \int d^Dx \, \Big ( F^\mu_a - \tilde J^\mu_a\Big  )
   \Big ( F^\nu_a -\tilde J^\nu_a\Big  )
   \Big ( F_{b\mu} -\tilde J_{b\mu} \Big  )
   \Big ( F_{b\nu} -\tilde J_{b\nu} \Big  ) \,
\label{app:inv}
\end{eqnarray}
In the above equation $D_\mu[F]$ stands for the covariant derivative w.r.t. $F_{a\mu}$
\begin{eqnarray}
D_\mu[F]_{ab}  = \delta_{ab} \partial_\mu + \epsilon_{acb} F_{c \mu} \,
\label{app:inv6}
\end{eqnarray}


%


\end{document}